\documentclass[5p,twocolumn,sort&compress]{elsarticle}
\usepackage{graphicx}
\usepackage{multirow}
\usepackage{amsmath,amssymb,amstext,amsthm,enumerate}
\usepackage[usenames,dvipsnames]{color}
\usepackage{tabularx}
\usepackage{url}
\usepackage{siunitx}
\graphicspath {{./figures/}}
\makeatletter
\def\input@path{{./figures/}}
\makeatother

\journal{Computer Physics Communications}

\begin{document}

\begin{frontmatter}
\title{GPU accelerated population annealing algorithm}
\author[Landau,SCC]{Lev Yu. Barash}
\author[Coventry]{Martin Weigel}
\author[SCC,PJSU]{Michal Borovsk\'{y}}
\author[Leipzig]{Wolfhard Janke}
\author[Landau,SCC,HSE]{Lev N. Shchur}
\address[Landau]{Landau Institute for Theoretical Physics, 142432 Chernogolovka, Russia}
\address[SCC]{Science Center in Chernogolovka,142432 Chernogolovka, Russia}
\address[Coventry]{Applied Mathematics Research Centre, Coventry University, Coventry, CV1 5FB, United Kingdom}
\address[PJSU]{P.J. \v{S}af\'{a}rik University, Park Angelinum 9, 040\,01 Ko\v{s}ice, Slovak Republic}
\address[Leipzig]{Institut f\"ur Theoretische Physik, Universit\"at Leipzig, Postfach 100920, 04009 Leipzig, Germany}
\address[HSE]{National Research University Higher School of Economics, 101000 Moscow, Russia}

\begin{abstract}
  
  Population annealing is a promising recent approach for Monte Carlo simulations in
  statistical physics, in particular for the simulation of systems with complex
  free-energy landscapes. It is a hybrid method, combining importance sampling
  through Markov chains with elements of sequential Monte Carlo in the form of
  population control. While it appears to provide algorithmic capabilities for the
  simulation of such systems that are roughly comparable to those of more established
  approaches such as parallel tempering, it is intrinsically much more suitable for
  massively parallel computing. Here, we tap into this structural advantage and
  present a highly optimized implementation of the population annealing algorithm on
  GPUs that promises speed-ups of several orders of magnitude as compared to a serial
  implementation on CPUs. While the sample code is for simulations of the 2D
  ferromagnetic Ising model, it should be easily adapted for simulations of other
  spin models, including disordered systems. Our code includes implementations of
  some advanced algorithmic features that have only recently been suggested, namely
  the automatic adaptation of temperature steps and a multi-histogram analysis of the
  data at different temperatures.

\end{abstract}
\end{frontmatter}

{\bf PROGRAM SUMMARY}

\begin{small}
\noindent
{\em Manuscript Title:} GPU accelerated population annealing algorithm\\
{\em Authors:} Lev Yu. Barash, Martin Weigel, Michal Borovsk\'{y}, Wolfhard Janke, Lev N. Shchur \\
{\em Program Title:} PAIsing \\
{\em Journal Reference:}                                      \\
  %Leave blank, supplied by Elsevier.
{\em Catalogue identifier:}                                   \\
  %Leave blank, supplied by Elsevier.
{\em Licensing provisions:}   Creative Commons Attribution license (CC BY 4.0)                                 \\
  %enter "none" if CPC non-profit use license is sufficient.
{\em Programming language:} C, CUDA                                \\
{\em Computer:} System with an NVIDIA CUDA enabled GPU \\
  %Computer(s) for which program has been designed.
{\em Operating system:}  Linux, Windows, MacOS   \\
  %Operating system(s) for which program has been designed.
{\em RAM:} 200 Mbytes   \\
  %RAM in bytes required to execute program with typical data.
{\em Number of processors used:} 1 GPU \\
  %If more than one processor.
{\em Supplementary material:}                                 \\
  % Fill in if necessary, otherwise leave out.
  % Please give some freely chosen keywords that we can use in a
  % cumulative keyword index.
{\em Classification:} 23   \\
  %Classify using CPC Program Library Subject Index, see (
  % http://cpc.cs.qub.ac.uk/subjectIndex/SUBJECT_index.html)
  %e.g. 4.4 Feynman diagrams, 5 Computer Algebra.
{\em External routines/libraries:} NVIDIA CUDA Toolkit 6.5 or newer     \\
  % Fill in if necessary, otherwise leave out.
{\em Subprograms used:}                                       \\
  %Fill in if necessary, otherwise leave out.
{\em Nature of problem: }%
The program calculates the internal energy, specific heat, several magnetization
moments, entropy and free energy of the 2D Ising model on square lattices of edge length $L$
with periodic boundary conditions as a function of inverse
temperature $\beta$.
% Describe the nature of the problem here.
\\
{\em Solution method: }%
The code uses population annealing, a hybrid method combining Markov chain updates
with population control. The code is implemented for NVIDIA GPUs using the CUDA
language and employs advanced techniques such as multi-spin coding, adaptive
temperature steps and multi-histogram reweighting.
% Describe the method solution here.
\\
{\em Restrictions: }%
The system size and size of the population of replicas
are limited depending on the memory of the GPU device used. \\
% Describe any restrictions on the complexity of the problem here.
{\em Unusual features:}\\
  %Describe any unusual features of the program/problem here.
{\em Additional comments:}\\
  %Provide any additional comments here.
Code repository at \url{https://github.com/LevBarash/PAising}.
{\em Running time:}  
For the default parameter values used in the sample programs, $L=64$, $\theta=100$,
$\beta_0=0$, $\beta_f=1$, $\Delta\beta=0.005$, $R=20\,000$, a typical run time on an
NVIDIA Tesla K80 GPU is 151 seconds for the single spin coded (SSC) and 17 seconds
for the multi-spin coded (MSC) program (see Sec.~\ref{sec:algorithm} for
a description of these parameters).
% Give an indication of the typical running time here.
\\

\end{small}

\section{Introduction}

Monte Carlo methods are among the core techniques for studying the statics and
dynamics of particle systems in classical and quantum physics, in particular for
systems in statistical physics \cite{binder:book2}. Although for a few problems
simple sampling is reasonably efficient, most applications are based on importance
sampling techniques. Among them, Markov chain Monte Carlo (MCMC) is by far the most
widely used approach in statistical physics. In quantum Monte Carlo, on the other
hand, one can evaluate the wave function in a path-integral formulation in imaginary
time by a swarm of particles diffusing in configuration space that undergo a sequence
of birth-death processes \cite{kosztin:96}. This is a special case of a procedure
more generally known as sequential Monte Carlo \cite{doucet:13}. Such procedures have
been more reluctantly adopted in statistical physics applications, but they have
gained some traction recently, for example in variants of ``go with the winners''
simulations \cite{grassberger:02a}. In sequential Monte Carlo, configurations are
gradually built in possibly biased steps, sequentially accumulating weights that
multiply configurations in the final averages. In many applications such weights
fluctuate wildly, thus leading to rather unstable results. In the ``go with the
winners'' approaches, configurations are selectively cloned or pruned in accordance
with their weight to tame these fluctuations, a procedure often referred to as
population control.

One recent approach of this type is the population annealing (PA) algorithm
\cite{iba:01,hukushima:03}. There, a large number of configurations are prepared in
independent equilibrium configurations, for instance at infinite temperature. Each
configuration evolves according to a standard MCMC approach at the given temperature.
The population is then gradually cooled, and each configuration sequentially builds
up a weight depending on its energy at the instant of temperature change. Population
control is used to keep weight fluctuations under control. We here focus on a variant
where ``perfect'' population control is used at each temperature step such that all
weights remain equal to unity at all times \cite{machta:10a}. The approach has been
successfully used for equilibrium simulations of spin-glass systems
\cite{wang:14,wang:15a,wang:15b}, and also for finding ground states in spin glasses
and other systems with frustrated interactions \cite{wang:15,borovsky:16}. Recently,
we have studied the behavior of PA for simulations of the 2D Ising model, analyzing
systematically the dependence on the population size and annealing protocol, and
proposed a number of improvements \cite{weigel:16}.

The era of serial computing came to an end in the early 2000s when CPU clock
frequencies first hit the ``wall'' of about 3.5 GHz, beyond which heat dissipation
becomes unfeasible with conventional techniques and the power consumption increases
too steeply. While Moore's law \cite{moore:65} predicting an exponential growth of
the number of transistors in an integrated circuit continues to hold, the resulting
exponential growth of computational power seen for CPUs essentially stopped being an
increase of serial performance (for instance through the increase of clock speeds)
and now translates into a corresponding increase in the number of parallel cores or
other compute units. Thus the comfortable situation where the same old code or, at
least, the same old algorithm could be run on more modern hardware with exponentially
decreasing run times with every new generation of machine, has come to an
end. Instead, it has become necessary to design and implement solutions to our
computational problems that scale well up to thousands or maybe even millions of
cores \cite{asanovic:06}. A computational environment that recently proved to be a
particularly useful pathway towards massively parallel computing are graphics
processing units (GPUs) and similar accelerator devices. They feature a much higher
density of actual compute units than CPUs, at the expense of reduced cache memories
and control logic units that are mostly useful for accelerating serial and
unpredictable loads, and are hence very well suited for the needs of scientific
computing \cite{owens:08,hwu:11}. In statistical physics, significant speed-ups have
been observed for Ising model simulations with local \cite{preis:09,weigel:10c} as
well as non-local \cite{weigel:10b} udpate algorithms; for continuous-spin systems
\cite{weigel:10a,yavorskii:12}; for spin glasses \cite{weigel:10a,lulli:15} and
random-field models \cite{navarro:16}; for Potts systems \cite{ferrero:11}; for
polymers \cite{gross:11} and many other applications.

The PA algorithm that requires the parallel simulation of a population of tens of
thousands up to millions of replicas appears to be a perfect match for this new type
of computational resource. The quality of approximation increases with population
size \cite{wang:15a,weigel:16} such that a higher parallel load is clearly
advantageous. As we will show below, we observe a GPU speed-up of around 230 times
over a serial CPU based code, thus bringing the wall-clock time for typical
calculations of the 2D Ising system considered here down to minutes in many
cases. For such models with Ising spins, the additional application of multi-spin
coding yields a further up to 10-fold speed-up, such that we reach a peak performance
of 10 ps per spin flip of the whole PA simulation code, including the resampling and
measurement parts. We provide a flexible implementation that can be configured using
command-line switches and should be easily adaptable to simulations of related models
such as 3D Ising systems, Potts and $O(n)$ models, and spin glasses. In extension to
the standard PA algorithm, our code also allows for the adaptive choice of inverse
temperature steps and an analysis of the simulation results with a multi-histogram
approach.

The rest of the paper is organized as follows. In Sec.~\ref{sec:algorithm} we
summarize the PA algorithm and the extensions employed here. Section
\ref{sec:implementation} discusses our implementation on GPU, while
Sec.~\ref{sec:msc} introduces the program variant that employs multi-spin coding. In
Sec.~\ref{sec:performance} we investigate the performance and reliability of our
code. Finally, Sec.~\ref{sec:conclusions} contains our conclusions.

\section{Algorithm}
\label{sec:algorithm}

The population annealing method was first discussed by Iba \cite{iba:01} in the
general context of population-based algorithms and later applied to spin glasses by
Hukushima and Iba \cite{hukushima:03}. More recently, Machta \cite{machta:10a} used a
method that avoids the recording of weight functions through population control in
every step. This is the variant we discuss and implement here.

\subsection{Population annealing}
\label{sec:main-algorithm}

As outlined above, the approach is a hybrid of sequential algorithm and MCMC that
simulates a population of configurations at each time, updating them with MCMC
methods and resampling the population periodically as the temperature is gradually
lowered. The algorithm can be summarized as follows:
\begin{enumerate}
\item Set up an equilibrium ensemble of $R_0 = R$ independent copies (replicas) of
  the system at inverse temperature $\beta_0$. Typically $\beta_0 = 0$, where this
  can be easily achieved.
\item To create an approximately equilibrated sample at $\beta_i > \beta_{i-1}$,
  resample configurations $j = 1,\ldots, R_{i-1}$ with their relative Boltzmann
  weight $\tau_i(E_j) = \exp[-(\beta_i-\beta_{i-1})E_j]/Q_i$, where
  \begin{equation}
    Q_i \equiv Q(\beta_{i-1},\beta_i) = \frac{1}{R_{i-1}}
    \sum_{j=1}^{R_{i-1}} \exp[-(\beta_i-\beta_{i-1})E_j].
    \label{eq:Q}
  \end{equation}
\item Update each replica by $\theta$ rounds of an MCMC algorithm at inverse
  temperature $\beta_i$.
\item Calculate estimates for observable quantities ${\cal O}$ as population averages
  $\sum_j {\cal O}_j/R_i$.
\item Goto step 2 unless the target inverse temperature $\beta_\mathrm{f}$ has been reached.
\end{enumerate}
If we choose $\beta_0 = 0$, equilibrium configurations for the replicas can be generated
by simple sampling, i.e., by assigning independent, purely random spin configurations
to each copy. The resampling process in step 2 can be realized in different ways
\cite{hukushima:03,machta:10a}. Here we use the following approach
\cite{wang:15}. For each replica $j$ in the population at inverse temperature
$\beta_{i-1}$ we draw a random number $r$ uniformly in $[0,1)$. The number of copies
of replica $j$ in the new population is then taken to be
\begin{equation}
  r_i^j = \left\{
    \begin{array}{ll}
      \lfloor  \hat{\tau}_i(E_j) \rfloor & \mathrm{if}\;\;r >  \hat{\tau}_i(E_j) - \lfloor
                                           \hat{\tau}_i(E_j) \rfloor \\
      \lfloor  \hat{\tau}_i(E_j) \rfloor + 1 & \mathrm{otherwise}
    \end{array}
  \right.,
  \label{eq:resampling-factors}
\end{equation}
where $\hat{\tau}_i(E_j) = (R/R_{i-1})\tau_i(E_j)$ is renormalized to ensure that the
population size stays close to the target value $R$. Here, $\lfloor x\rfloor$ denotes
the largest integer that is less than or equal to $x$ (i.e., rounding down). The new
population size is $R_i = \sum_j r_i^j$. This method requires only a single call to
the random number generator for each replica in the current population and leads to
very small fluctuations in the total population size. Note that it is possible that
$r_i^j = 0$, in which case the corresponding replica disappears from the population,
while other configurations will be replicated several times. In the standard setup,
steps of equal size in inverse temperature are taken, i.e.,
\[
\beta_i = \beta_{i-1} + \Delta\beta,
\]
and $\Delta\beta$ is an adjustable parameter. We discuss an adaptive, automatic
choice of temperature steps below in Sec.~\ref{sec:adaptive}. In the code presented
here, we use $\theta$ sweeps of Metropolis single-spin flip updates to equilibrate
each replica in each temperature step. Other updates such as heat-bath dynamics or
even non-local cluster moves could be employed easily as well.

Measurements are taken as population averages, and our code produces estimates for
the following quantities,
\begin{equation}
  \begin{split}
  \overline{e} &= \frac{1}{R_i} \sum_j E_j/N, \\
  C &= \beta^2 N \left(\overline{e^2}-\overline{e}^2\right), \\
  \overline{|m|} &=  \frac{1}{R_i} \sum_j |M_j|/N, \\
  \overline{m^2} &=  \frac{1}{R_i} \sum_j (M_j/N)^2, \\
  \overline{m^4} &=  \frac{1}{R_i} \sum_j (M_j/N)^4.
  \end{split}
  \label{eq:observables}
\end{equation}
Here, $E_j$ denotes the configurational energy and $M_j$ the configurational
magnetization of replica $j$, and $N$ is the number of spins. Additionally, PA
provides a natural estimate of the free energy,
\begin{equation}
  -\beta_i F({\beta_i}) = \ln Z_{\beta_0} + \sum_{k=1}^{i} \ln Q_k,
  \label{eq:free-energy}
\end{equation}
where $Z_{\beta_0}$ is the partition function at inverse temperature $\beta_0$,
$Z_{\beta_0} = 2^N$ for Ising spins and $\beta_0 = 0$, and $Q_k$ is the reweighting
factor \eqref{eq:Q} used in the resampling. From Eqs.~(\ref{eq:observables}) and
(\ref{eq:free-energy}) we can also compute the entropy per site via
\begin{equation}
  S(\beta_i)/N = \beta_i \overline{e}(\beta_i) - \beta_i F(\beta_i)/N. 
  \label{eq:entropy}
\end{equation}

\subsection{Weighted averages}
\label{sec:weighted}

It was shown in Ref.~\cite{machta:10a} that one of the strengths of the PA approach
is that by combining the data from independent runs not only statistical errors are
decreased, but also systematic deviations can be reduced. This is the case if one
uses weighted averages of results of independent runs. As was shown in
Refs.~\cite{machta:10a,wang:15a}, an unbiased way of combining the results of $M$
independent runs of PA for the same system and target population size is to weight
them by the free energies as estimated according to Eq.~(\ref{eq:free-energy}),
\begin{equation}
  \label{eq:weighted-average}
  \tilde{A}(\beta_i) = \sum_{m=1}^M \omega_m(\beta_i) \bar{A}_m(\beta_i),
\end{equation}
with
\begin{equation}
  \omega_m(\beta_i) = \frac{e^{-\beta_i F_m(\beta_i)}}{\sum_m e^{-\beta_i F_m(\beta_i)}},
\end{equation}
where $\bar{A}_m(\beta_i)$ denotes the average of observable $A$ in simulation $m$
and $F_m(\beta_i)$ the corresponding free-energy estimator according to
Eq.~(\ref{eq:free-energy}). The concept of weighted averages allows for an additional
parallelization in splitting the total simulation into independent parts. The
weighting ensures that this does not substantially degrade the quality of the results
\cite{wang:15a,weigel:16}. Note that the concept of weighted averages is more general
than the PA approach \cite{ferrenberg:89a}, but for the present algorithm the
necessary free-energy estimates are a free by-product of the simulation according to
Eq.~\eqref{eq:free-energy}. In the implementation provided here, multiple runs can be
requested on the command line, but the weighted averaging of results is left to the
user to perform separately.

\subsection{Adaptive temperature steps}
\label{sec:adaptive}

While an annealing cycle of the population is valid for any choice of the temperature
sequence $\beta_0$, $\beta_1$, $\ldots$, and given a sufficiently large number
$\theta$ of MCMC sweeps employed at each temperature it also leads to essentially
unbiased estimates of observables, the resampling step is only effective if
$\beta_i-\beta_{i-1}$ is sufficiently small \cite{weigel:16}. The optimal size of
temperature steps will itself depend on temperature, and a uniform stepping is not in
general ideal. As was recently shown in Ref.~\cite{weigel:16} uniform effectiveness
of resampling is achieved by ensuring a constant overlap of the energy histograms of
population members between the neighboring temperatures. This overlap can be computed
from the reweighting factors before actually performing the resampling step, and one
finds
\begin{equation}
  \alpha(\beta_{i-1},\beta') = \frac{1}{R_{i-1}}\sum_{j=1}^{R_{i-1}} \min\left(1,
    \frac{R\exp[-(\beta'-\beta_{i-1})E_j]}{R_{i-1}Q(\beta_{i-1},\beta')}
  \right).
\end{equation}
Clearly, $0\le \alpha(\beta_{i-1},\beta') \le 1$, and one can use numerical root
finding techniques such as, for instance, bisection search, to find $\beta'$ such
that $\alpha(\beta_{i-1},\beta') = \alpha^\ast$ and then set $\beta_i =
\beta'$.
Values of $0.5 \lesssim \alpha^\ast \lesssim 0.9$ provide sufficient histogram
overlap without an unnecessary proliferation of temperature steps. In practice, if
$M$ runs are performed for additional averaging, our code used in adaptive mode
decides about temperature steps only in the first run and keeps the temperature
sequence fixed for the remaining passes.

\subsection{Multi-histogram reweighting}
\label{sec:mhr}

As a PA sweep produces samples at a large number of closely spaced temperatures
(typically at least 100, even for small systems), it is natural to combine these data
to increase the accuracy and reduce statistical fluctuations in the spirit of the
multi-histogram analysis of Ferrenberg and Swendsen \cite{ferrenberg:89a}. Neglecting
correlations between the data at different temperatures as well as the effect of
autocorrelations, an optimized combination of histograms to yield an estimate of the
density of states is given by \cite{weigel:16}
\begin{equation}
  \Omega(E) = \frac{\sum\limits_{i = 1}^{N_\beta} P_{\beta_i}(E) }{\sum\limits_{i = 1}^{N_\beta} R_i
    \exp[\beta_i F(\beta_i)-\beta_iE]}.
  \label{eq:MHReq1}
\end{equation}
Here, $N_\beta$ denotes the total number of temperatures, and we assumed a
normalization of the histogram at inverse temperature $\beta_i$ such that
$\sum_E P_{\beta_i}(E) = R_i$. We note that the storage requirements are moderate as
at each time one only needs to store the sum of histograms up to the current
temperature and not each histogram individually. Generalizations to other quantities
such as magnetizations are possible \cite{weigel:16}.

\section{GPU realization}
\label{sec:implementation}

For definiteness we focus on an implementation for the ferromagnetic, zero-field
Ising model on the square lattice with Hamiltonian
\begin{equation}
  \label{eq:hamiltonian}
  {\cal H} = -J\sum_{\langle i,j\rangle}s_i s_j.
\end{equation}
Here, interactions are only between nearest neighbors $\langle i,j\rangle$ and
periodic boundary conditions are assumed. As is well known, this model undergoes a
continuous phase transition at the inverse temperature
$\beta_c = \frac{1}{2}\ln(1+\sqrt{2})$ \cite{mccoy:book}. The question of how well
suited population annealing is as a simulation technique to study this model and its
transition is discussed in Ref.~\cite{weigel:16}. Here, we are not concerned with
this aspect, but we use this model as a convenient starting point since a wealth of
exact or extremely accurate results are available for it as reference points, and a
generalization of the code to other spin models and even more general systems such as
polymers or particle systems should be rather straightforward.

\paragraph{General considerations}
Inspecting the algorithm given in Sec.~\ref{sec:main-algorithm}, one identifies three
computationally demanding steps: a resampling of the population that involves the
determination of weight factors and the copying of replicas, the update of individual
configurations with MCMC moves (i.e., spin flips), and the measurement of
observables. As we shall see below when reporting the performance results, most time
(on CPU or GPU) is normally spent on spin flips (see also Ref.~\cite{weigel:16}),
while for typical choices of $\theta$ ($\theta \ge 10$, say) the resampling step and
the measurements of the elementary quantities listed in
Eqs.~\eqref{eq:observables}--\eqref{eq:entropy} are much less time
consuming\footnote{As we shall see below, however, this balance is somewhat changed
  for the case of a multi-spin coded implementation.}. These observations suggest to
also choose the effort for optimization of each of these parts correspondingly. We
hence first focus on the spin updates.

One of the basic features of present day GPU devices that is of paramount importance
for performance is the technique of {\em latency hiding\/} implemented in the
scheduling algorithm \cite{kirk:10}. Each time an elementary group of threads (given
by a warp of 32 threads on current NVIDIA GPUs) accesses some data in memory that is
currently not cached, there is a latency of hundreds or even thousands of clock
cycles until the read or write operation completes. Instead of leaving the compute
units idle, the scheduler puts the present warp in the queue and allows another warp
that has already completed its data transaction to use the compute units. If only
enough such thread groups are available, the compute cores will be kept constantly
busy and hence the memory latencies are hidden away. Good GPU performance thus
requires to break the work into many threads, optimal performance is often only
reached for thread numbers in excess of ten times the number of available physical
cores \cite{gross:17}.

A second crucial requirement for exploiting the full potential of GPUs relates to the
minimization of costly global memory accesses. This includes a reasonable level of
compression of the data to be transferred for the updates. For the present problem
with Ising variables $s_i = \pm 1$ it suggests to use the narrowest available native
data type to represent spins, which is an 8-bit integer, or to revert to a multi-spin
coding approach. A discussion of the latter technique is postponed until the next
section. Further, the relative slowness of memory makes it useful to cache and re-use
data as much as possible, which could involve using automatic caches or the
user-managed cache known as shared memory \cite{kirk:10}. Finally, it implies
optimization geared towards increasing the locality of memory operations as each
direct access to global memory (implying a cache miss) fetches a full cache line of
128 bytes. Ideally, the threads in a warp access memory locations in the same cache
line(s), thus making use of all of the data that is actually loaded. This concept is
known as {\em memory coalescence\/}.

\paragraph{Spin updates}
By construction population annealing suggests to parallelize the calculations for
different members of the population. One particularly simple code setup is hence to
assign one thread to the updating of each replica such that in total $R_i$ threads
are used for the MCMC part of the algorithm, i.e., for flipping spins. Each thread
then goes sequentially through the lattice. To ensure good memory coalescence in this
case, the same spins of each replica should be placed next to each other in memory,
so configurations should be stored in replica-major order. In practice this code
setup, which we denote as {\em replica-parallel\/}, does show good but not optimal
performance, especially for smaller population sizes where it does not provide enough
parallelism. Where this approach does not provide optimal performance, it still has
the advantage of being completely general, and it could consequently be applied
unaltered to PA simulations of any other model. We hence mention it here as a safe
fall-back solution especially for problems for which it is not possible or
straightforward to implement a domain decomposition (for instance for systems with
long-ranged interactions).

\begin{figure}[tb]
  \centering
  \includegraphics[width=0.95\columnwidth]{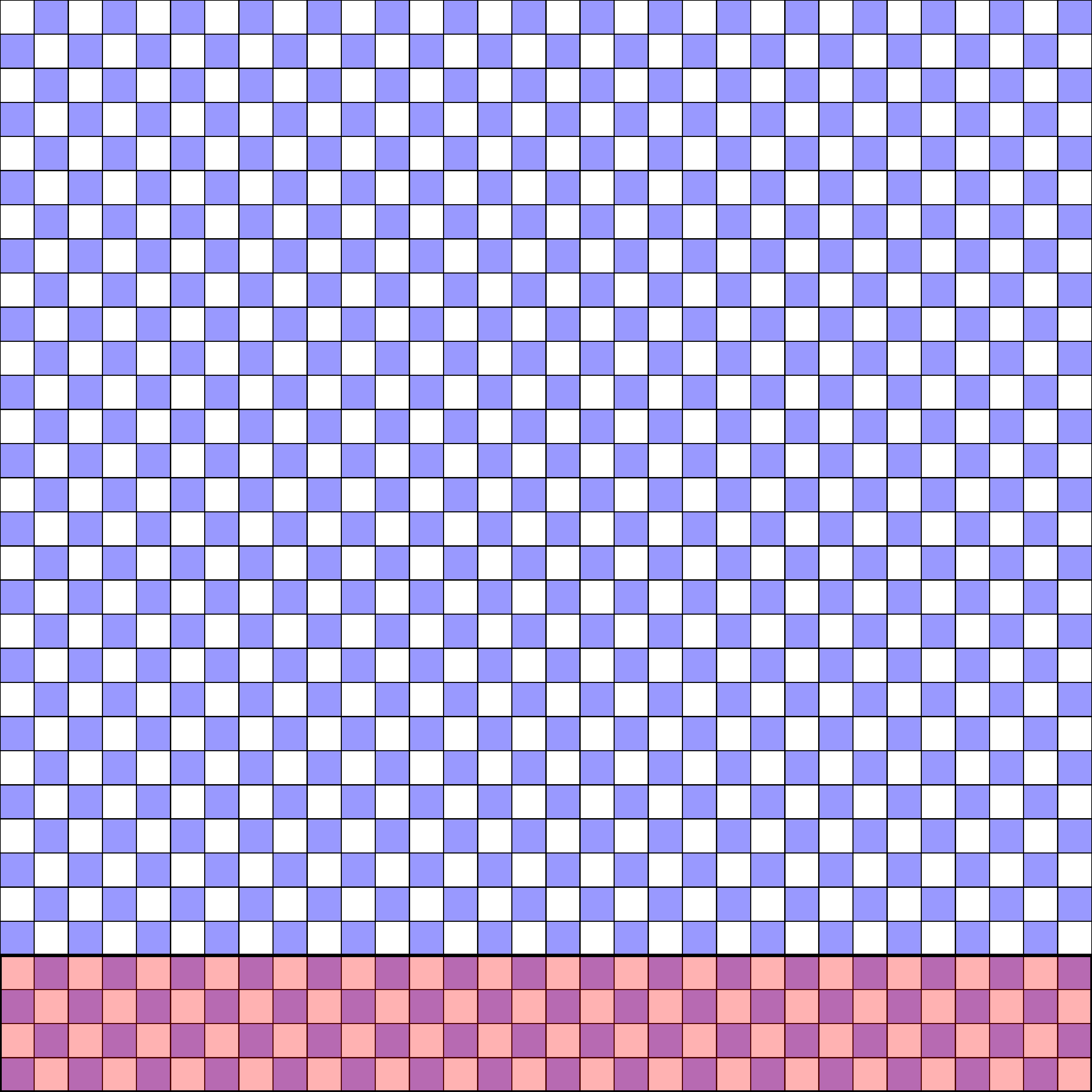}
  \caption{Diagrammatic representation of the mapping of thread blocks to spins in
    the updating kernel. The code works with thread blocks of size
    \texttt{EQthreads}. Each block works on a single replica of
    the population, using its threads to update tiles of size
    $2\times\mathtt{EQthreads}$ spins. To this end it flips spins on one
    checkerboard sub-lattice first, moving the tiles over the lattice until it is
    covered, sychronizes and then updates the other sub-lattice.  }
  \label{fig:checker}
\end{figure}

To increase the amount of parallel work, for the present code for the 2D Ising model
we opted to additionally parallelize the updates for each single replica, using a
domain decomposition of the lattice. This was extensively used previously for
simulations using MCMC (single-spin flips) only. The basic step consists of a
checkerboard decomposition of the lattice which allows for independent updates of all
spins of one sub-lattice\footnote{Generalizations to other lattice structures and
  larger, but finite interaction ranges are possible \cite{weigel:10a}.}. We denote
the corresponding scheme that parallelizes over replicas and spins as {\em
  spin-parallel\/}. As the number of threads per block is limited to 1024 on current
CUDA devices, one either lets each thread update a certain range of spins or employs
a further decomposition of the lattice, be it in strips \cite{preis:09}, a second
layer of checkerboard tiles \cite{weigel:10a,weigel:10c} or some other form of
subdivision \cite{lulli:15}. For the present code, we used one of the simplest
solutions and let the \texttt{EQthreads} threads of a block employed for the
spin-updating kernel \texttt{checkKerALL()} handle the spins of a full replica in the
following way (cf.\ Fig.~\ref{fig:checker}): the first \texttt{EQthreads} spins of
the blue sub-lattice are updated in parallel, then the next \texttt{EQthreads} blue
spins and so forth until all blue spins of the replica have been dealt with. After a
synchronization of all threads they update the white spins of the current
configuration in the same way, followed by another synchronization of
threads. Finally, this whole procedure is repeated $\theta$ times until all spin
updates have been implemented. This setup is illustrated in Fig.~\ref{fig:checker}
for an $L=32$ lattice and $\mathtt{EQthreads} = 64$. To increase memory coalescence
we store the spins of each sub-lattice together, separate from the spins of the other
sub-lattice. Note that for this setup the spins are stored in a {\em spin-major\/}
order as the threads of a block work on spins in the same replica. We are not
explicitly using shared memory for the spin flips as it was not found to improve
performance on the devices tested here. A further optimization could consist of
storing the spin arrays in texture memory as suggested in Ref.~\cite{lulli:15} which
simplifies index arithmetics and allows to make use of the separate texture cache,
but for the sake of simplicity we refrain from such additional optimizations that are
expected to yield only quite moderate further speed improvements.

In this spin-parallel setup utilizing additional parallel work inside of each
replica, different replicas are handled by different thread blocks. We request $R_i$
thread blocks at kernel invocation which will cause no problems for realistic
population sizes on recent devices where the maximum number of blocks is
$2^{31}-1 \approx 2\times 10^9$.  For the actual spin updates, we use pre-calculated
tables of the Metropolis factors $\exp(-\beta\Delta E)$, stored in texture
memory\footnote{Note, however, that these tables need to be re-calculated for each
  (inverse) temperature step.}  \cite{weigel:10a}. For deciding about the acceptance
of proposed spin flips the algorithm requires one random number per spin
update. Random number generation on GPUs and in other massively parallel contexts
requires a way of producing many uncorrelated (sub-)\-sequences, and certain
parameters such as the memory footprint make some of the standard generators in
serial environments unsuitable for a massively parallel application. Some of the
related issues are discussed in Refs.~\cite{manssen:12,barash:14}. A suite of
generators loosely based on cryptographic algorithms turned out to be particularly
competitive in this context, namely the series of Philox generators of
Ref.~\cite{salmon:11}. In the tests conducted in Ref.~\cite{manssen:12} it combined
excellent GPU performance with a passing of all tests of the TestU01 suite
\cite{lecuyer:07}. Also, in the meantime it has been included as one of the
generators in the \texttt{curand} library that is part of NVIDIA's CUDA
distribution. It hence requires no further code to be used for the present
application. Additionally, users can readily replace it by any of the alternative
RNGs included in \texttt{curand} if they so desire. One of the important advantages
of the Philox generator is that it does not require the transfer of a generator state
between GPU main memory and the multiprocessors doing the actual calculations. This
is a consequence of it being a counter-based generator, i.e., the generation of the
number $x_n = f(n)$ in the sequence does not require knowledge of $x_{n-1}$ or any
other previous state. We use one instance of the Philox\_4x32\_10 generator per
thread, which is initialized in the kernel with a sequence number determined from the
grid and block indices of the thread and a global iteration parameter. The required
numbers in $(0,1)$ are then generated by in-line calls to \texttt{curand\_uniform()}
in the spin-updating kernel \texttt{checkKerALL()}. This inline production of random
numbers is faster than a pre-generation in dedicated arrays in a separate kernel and
also much more efficient in terms of the memory footprint as no arrays are required.

\paragraph{Resampling and measurements}
The resampling process is also fully handled on GPU. To determine the resampling
factors $\hat{\tau}_i(E_j)$ of Eq.~\eqref{eq:resampling-factors} one first needs the
normalization constant $Q_i$ of Eq.~\eqref{eq:Q}, which equals the sum of all
(unnormalized) resampling factors. There are $R_i$ summands, and the corresponding
kernel \texttt{QKer()} is called with $\mathtt{Nthreads}$ threads best chosen to
equal the maximum block size (1024 for current NVIDIA GPUs) and, correspondingly,
$\lceil R_i/\mathtt{Nthreads}\rceil$ blocks. (Here, $\lceil x\rceil$ denotes the
smallest integer that is larger or equal to $x$, i.e., rounding up.) Within each
block, we use the standard parallel reduction method that adds elements pairwise in
several generations until only one element (the sum) is left --- a scheme that can be
visualized as a binary tree \cite{mccool:12}. This approach would typically store the
intermediate results in shared memory. On devices that support it, however, we find
it to be faster to use the ``shuffle'' operations that allow threads to access
registers from different threads in the same warp\footnote{For the case of devices
  with compute capability $\le 3.0$, where shuffle operations are not available, we
  revert to a reduction with partial results stored in shared memory.}. As threads
from different blocks cannot directly communicate, the sum of partial results of each
thread block is typically determined by an additional kernel invocation
\cite{kirk:10}.  Alternatively, one can make use of the \texttt{atomicAdd()} device
function provided by CUDA to complete the reduction in the same kernel call.  Since
for the solution using \texttt{atomicAdd()} the order of summation is not well
defined, different runs with the same parameters and random number seeds could
potentially lead to slightly different values of $Q_i$ (at the level of the
floating-point precision). As this enters the resampling part of the code, where the
realized number of copies depends on the comparison of a quantity involving $Q_i$ to
a random number according to Eq.~\eqref{eq:resampling-factors}, we cannot exclude
that the outcome depends on the unpredictable order of atomic operations in some
marginal cases. To have a deterministic code that provably simulates the same set of
configurations in each run with the same parameters (including the random-number
seed), we decided to sum the per-block partial results for $Q_i$ in a subsequent
kernel call, thus making this part deterministic. For the calculation of averages
discussed below, we use the semantically simpler code with \texttt{atomicAdd()}. A
second kernel, \texttt{CalcTauKer()} is used with the same execution configuration to
determine the number of copies of each replica to be created according to
Eq.~\eqref{eq:resampling-factors}. Here, another random number is used for each
replica in the current population to determine whether the number of copies is
$\lfloor \hat{\tau}_i(E_j) \rfloor$ or $\lfloor \hat{\tau}_i(E_j) \rfloor+1$. To
facilitate the parallel placement of new copies in the vector storing the resampled
population, we also calculate the partial sums $\sum_{j=1}^k r_i^j$, i.e., the
offsets into that vector, again using the same parallel reduction approach. This
calculation is completed in the kernel \texttt{CalcParSum()}. In the end,
\texttt{resampleKer()} is used to copy the selected individual replicas into the
previously calculated locations of the new population vector, using one thread per
spin in a tile of size $\mathtt{EQthreads}$ and a number of blocks that covers the
full population and each individual lattice with tiles.

Finally, measurements of the quantities of
Eqs.~\eqref{eq:observables}--\eqref{eq:entropy} are computed using a parallel
reduction algorithm to first calculcate the configurational energy and magnetization
of each replica in the kernel \texttt{energyKer()}. As only one block is assigned to
each replica in this case, no further reduction of block values is required
here. Finally, another parallel reduction is used in the kernel
\texttt{CalcAverages()} to determine the population averages, employing
\texttt{atomicAdd()} for the inter-block reduction.

\paragraph{Further optimization and parameters}
We note that through the fluctuations of $R_i$ the execution configuration of the
kernels changes with each temperature step. For a fully loaded GPU this causes only
negligible variations in the total performance, however. An important feature of the
provided implementation is that all calculations are performed on GPU, so no
significant memory transfers to or from CPU occur during the PA run time. A number of
further optimizations have been employed to achieve good performance. We request a
larger L1 cache over shared memory using the \texttt{cudaDeviceSetCacheConfig()}
command as this turns out to be beneficial for the memory accesses in the main
\texttt{checkKerALL()} kernel that does not make use of shared memory. There is a
maximal number of threads that can be resident on a multiprocessor at any given time,
and in general it is found that latency hiding works better the more threads are
resident. This {\em occupancy\/} of multiprocessors can be limited by the number of
available registers, however. Depending on the GPU employed, it can be beneficial to
request a maximum number of registers to be consumed per thread using the
command-line option \texttt{--maxregcount} of the \texttt{nvcc} compiler. The
occupancy achieved with a given setup and register usage can be determined using the
occupancy calculator spreadsheet that comes with the CUDA distribution.

We provide here two separate codes, one for single-spin coding and one for multi-spin
coding (see below). The relevant parameters such as $R$, $\theta$, $\beta_0$ and
$\beta_\mathrm{f}$, as well as the number of runs $M$ can be specified either through
constants (\texttt{\#define}s) at the beginning of the source code or through
command-line arguments. There are two GPU specific parameters, \texttt{EQthreads} and
\texttt{Nthreads}, which decide about the block size in the different kernels. These
can be adjusted by changing the values in the \texttt{\#define}s in the source code,
but the default choices, $\mathtt{EQthreads} = 128$ and $\mathtt{Nthreads} = 1024$,
are virtually always (near) optimal on modern cards. The seed of the RNG can be
changed by adapting \texttt{RNGseed}, and in the default setup it is initialized
using the system time.

The results of each run of the algorithm are stored in a separate output file in text
format. Each line of the output contains the values $\beta$, $\overline{e}$, $C$,
$\overline{m}$, $\overline{m^2}$, $\overline{m^4}$, $\beta F/N$, $S/N$, $R$, $\ln Q$,
i.e., the inverse temperature, energy per site, specific heat, magnetization per site
and its moments, free energy density divided by temperature, entropy per site,
population size, and logarithm of partition function ratio, respectively.

\section{Multi-spin coding} 
\label{sec:msc}

It is clear from the general design principles for efficient GPU code as discussed
above in Sec.~\ref{sec:implementation} that a minimization of memory transfers will
often result in more efficient code. More specifically, this will always be the case
for code that is {\em memory-bound\/}, i.e., for which the mix of memory transactions
and arithmetic operations is such that the performance limiting factor is the latency
and bandwidth of memory transactions \cite{kirk:10}. Since the Metropolis update of
the Ising model used in the equilibrating subroutine is arithmetically very light,
especially when using a precomputed table for the exponential function, this is
indeed the case for the present application. Under these circumstances any
modification that reduces memory transfers can be expected to increase
performance. As an Ising spin is a single-bit variable, it is clear that storing
it in a standard built-in variable (even if it is of 8-bit length) is not ideal and
an explicit one-bit representation promises some performance improvement. This can be
implemented using multi-spin coding (MSC), i.e., by storing the states of $p$ spin
variables in a single machine word of $p$ bits \cite{zorn:81}. Natural choices for
the architecture are $p=8$, $16$, $32$ and $64$. While for simulations of single
systems as discussed in Refs.~\cite{zorn:81,ito:90} the spins represented by $p$ bits
in a word correspond to different lattice sites (synchronous MSC), for the present
application it is more convenient to have the bits in a word represent the spins on
the same lattice site but in different replicas (asynchronous MSC)
\cite{belleti:09,weigel:10a}. Quite efficient bitwise operations are available to
implement a parallel Metropolis update of the spins coded in a $p$-bit word. This
approach has been extensively used in simulations, in particular, of spin-glass
models \cite{hasenbusch:08,belleti:09,fang:14,lulli:15,fernandez:16}.

\begin{table}[tb]
  \caption{%
    Speedup of MSC implementations of PA with $p$ spins per word as compared to the SSC version.  The
    calculations were performed on a Tesla K80 card, but very similar results are
    expected for other GPUs. The parameters of the simulations were $L=128$, $\theta=500$,
    $R=80\,000$, and $\Delta\beta = 0.02$.
    The spin update used $n_\mathrm{RNG}$ random numbers of
    the underlying generator (Philox) to decide about the acceptance of flips of the
    $p$ spins coded in a word. For the data in the last section an additional linear
    congruential generator seeded by the underlying generator (Philox) is used to
    generate $p$ derived random numbers for the flipping of
    individual spins. The last column indicates the increase in statistical errors
    (in the low-temperature phase) through the re-use of random numbers
    [see also Fig.~\ref{fig:msc}(c)].
  }
  \label{tab:msc}
  \begin{center}
  \begin{tabular}{|r|r|c|S[table-format=2.2]|S[table-format=2.2]|}
    \hline
    $p$ & $n_\mathrm{RNG}$ & LCG & \multicolumn{1}{c|}{speedup} & 
    \multicolumn{1}{c|}{$\sigma_\mathrm{MSC}^2(C)/\sigma_\mathrm{SSC}^2(C)$} \\
    \hline
8  & 8                        & NO                  & 1.76 &  1 \\
16  & 16                       & NO                  & 1.73 & 1 \\
32  & 32                       & NO                  & 1.77 & 1 \\
64  & 64                       & NO                  & 1.72 & 1 \\
\hline
8  & 1                        & NO                  & 6.53 & 8  \\
16  & 1                       & NO                  & 10.44 & 16 \\
32  & 1                       & NO                  & 13.33 & 32 \\
64  & 1                       & NO                  & 9.23 & 64  \\
\hline
8  & 1                        & YES                  & 6.00 & 1 \\
16  & 1                       & YES                  & 8.73 & 1 \\
32  & 1                       & YES                  & 9.95 & 1 \\
64  & 1                       & YES                  & 7.92 & 1 \\
\hline
  \end{tabular}
  \end{center}
\end{table}

\begin{figure}[tb!]
  \begin{center}
    \includegraphics[width=0.95\columnwidth]{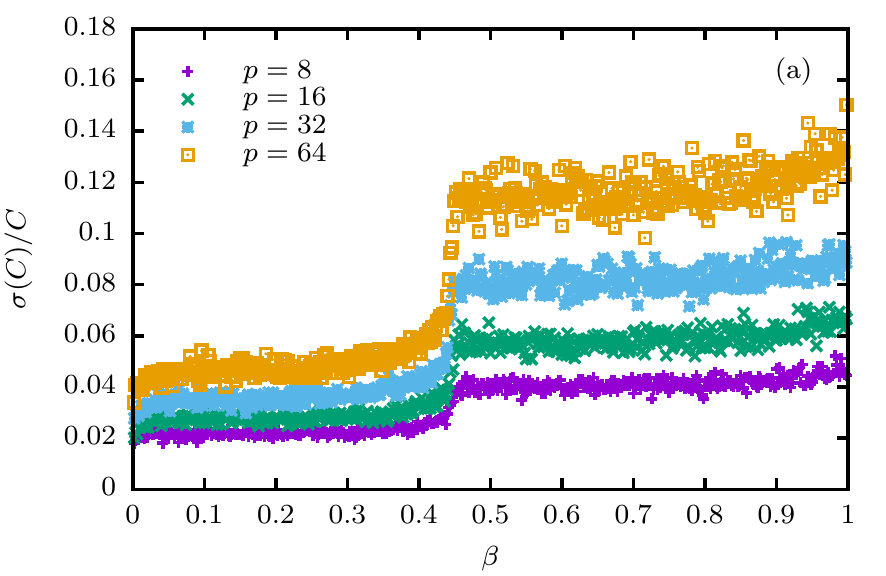}
    \includegraphics[width=0.95\columnwidth]{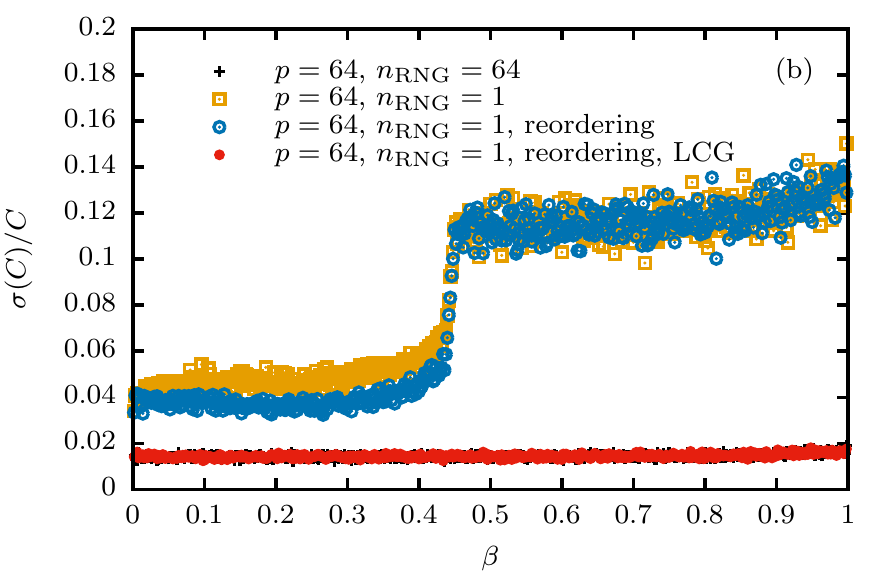}
    \includegraphics[width=0.95\columnwidth]{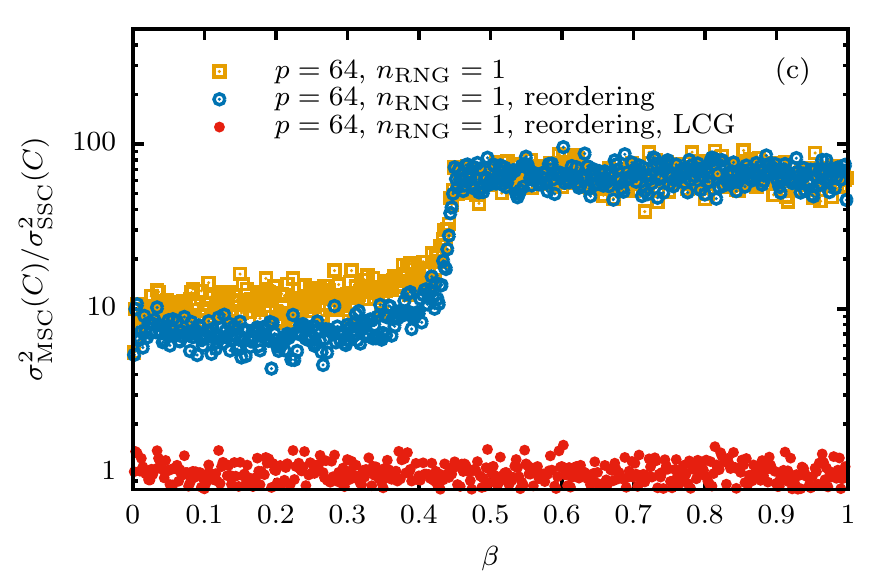}
    \caption{%
      (a) Relative error of estimates of the specific heat from PA runs for an $L=64$
      2D Ising system with $\theta=100$, $\Delta\beta = 0.002$ and population size
      $R=10\,000$ using asynchronous multi-spin coding (MSC) with $p=8$, $16$, $32$ and $64$
      bits. The same random numbers are used to decide about spin flips in all replicas
      coded in the same words. (b) Relative error of the specific heat for $p=64$ with
      $n_\mathrm{RNG} = 64$ and for $n_\mathrm{RNG} = 1$ with additional reordering of
      replicas and with using different random numbers for each replica coded in the same
      word, produced by an in-line linear congruential generator (LCG) seeded by the main
      generator (Philox). The data for this last variant and that of $n_\mathrm{RNG}
      = 64$ are practically indistinguishable. (c) Ratio of the estimated variances
      $\sigma^2(C)$ of the multi-spin coded (MSC) variants as compared to the
      single-spin coded (SSC) reference implementation.
    }
    \label{fig:msc}
  \end{center}
\end {figure}

The resulting MSC variant of the code shows increased performance over a single-spin
coded (SSC) version, with an improvement that depends only weakly on the number of
spins $p$ coded together. This is illustrated in the data in the first section of
Table~\ref{tab:msc}, where a different random number is drawn using the base
generator for each of the $p$ spins coded in a word, i.e., $n_\mathrm{RNG}=p$. The
relatively moderate and mostly $p$ independent improvement is a result of the fact
that the time taken per spin update is in this setup limited by the time it takes to
generate the random numbers used to decide about the acceptance of spin flips. A
number of implementations of this scheme for spin glasses
\cite{hasenbusch:08,fang:14,lulli:15} have used the {\em same\/} random number for
deciding about flipping all of the $p$ spins in a word. This introduces some
correlations, however, and while it is argued that this effect is minor for
spin-glass problems due to the property of bond chaos in such systems \cite{bray:87},
we expect it to be much more relevant for the case of the ferromagnet studied
here. If the same random numbers are used for deciding about flips of the $p$ spins
coded together, this implies that these replicas develop in a correlated manner. In
particular, if (some of) these $p$ replicas have identical spin configurations as is
the case if they are copies of the same parent configuration in the resampling
process, they are coupled and remain identical for all future times. This clearly
interferes with the goal of fair sampling. To illustrate this effect, we show in
Fig.~\ref{fig:msc}(a) the relative errors of the specific heat of the 2D Ising model
sampled with the MSC PA implementation with $p=8$, $16$, $32$ and $64$ spins coded
together, respectively, while using the same random numbers to flip spins in all $p$
replicas coded together. As is clearly seen, the errors in this setup increase with
$p$, and a rescaling of the $y$ axis reveals that $\sigma(C)/C$ in fact increases
proportional to $\sqrt{p}$ as expected from general statistical arguments (not
shown). On the other hand, the performance of this variant using only one random
number for $p$ spins is found to be excellent, cf.\ the data in the second section of
Table~\ref{tab:msc}. Note that here in contrast to the case with $n_\mathrm{RNG} = p$
the speedup varies considerably with $p$, and we find the best result for $p=32$.

In an attempt to alleviate the correlation effect, we introduced a rearrangement of
replicas after resampling in such a way as to avoid placing offspring of the same
parent configuration in the same word. This is achieved by the following procedure.
If we enumerate all replicas as $k = 1$, $\ldots,$ $R_i$, then for a given $n$,
$0\le n < \lceil R_i/p\rceil$, the spins of the replicas $np+1$, $np+2$, $\dots$,
$np+p$ are originally stored in the same words. Or, equivalently, replicas with the
same value of $\lceil k/p\rceil$ are stored in the same word. The population is then
rearranged such that replicas with the same value of $k \bmod \lceil R_i/p \rceil$
are stored in the same word, i.e., when initially replicas $1$, $2$, $\ldots$, $p$
occupy the first word, this now contains replicas $1$, $\lceil R_i/p\rceil+1$,
$\ldots$, $(p-1)\lceil R_i/p\rceil+1$.  This process can be pictured as transposing a
$p\times\lceil R_i/p\rceil$ matrix, followed by reshaping the result to again occupy
$p$ rows. Unless a parent has more than $\lceil R_i/p \rceil$ children (which is
unlikely for sufficiently large populations), this setup ensures that descendants
from the same parent configuration are placed in different words. As a result, their
next spin updates are governed by independent random number samples. However, it is
clear that at a lower temperature some of these sibling replicas could again end up
in the same machine word and hence remain correlated. The behavior of statistical
errors of the resulting improved algorithm is illustrated in
Fig.~\ref{fig:msc}(b). It leads to a slight reduction of the inflation of statistical
errors against the non-MSC implementation, but by no means removes it. Additionally,
the improvement appears to vanish for temperatures below the transition point
$\beta_c = \ln(1+\sqrt{2})/2$.

\begin{figure}[tb!]
  \begin{center}
    \includegraphics[width=0.95\columnwidth]{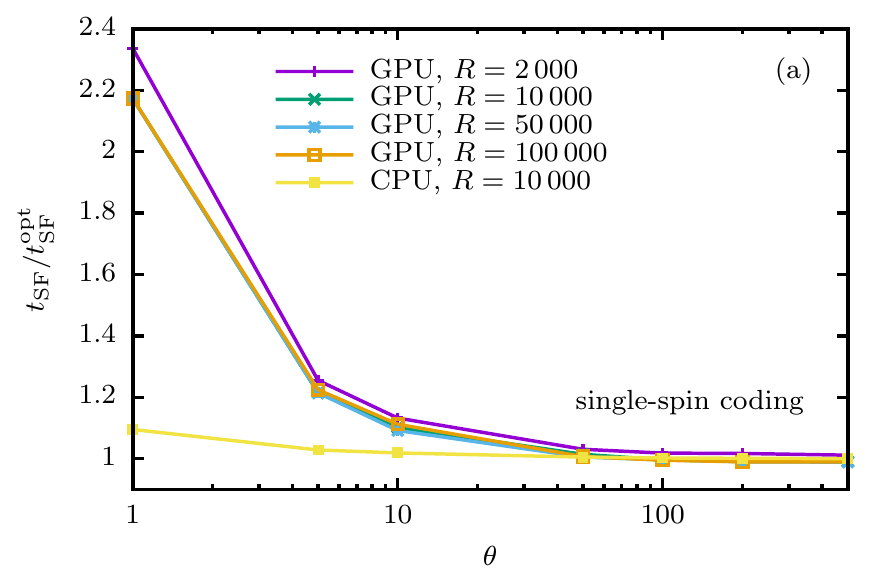}
    \includegraphics[width=0.95\columnwidth]{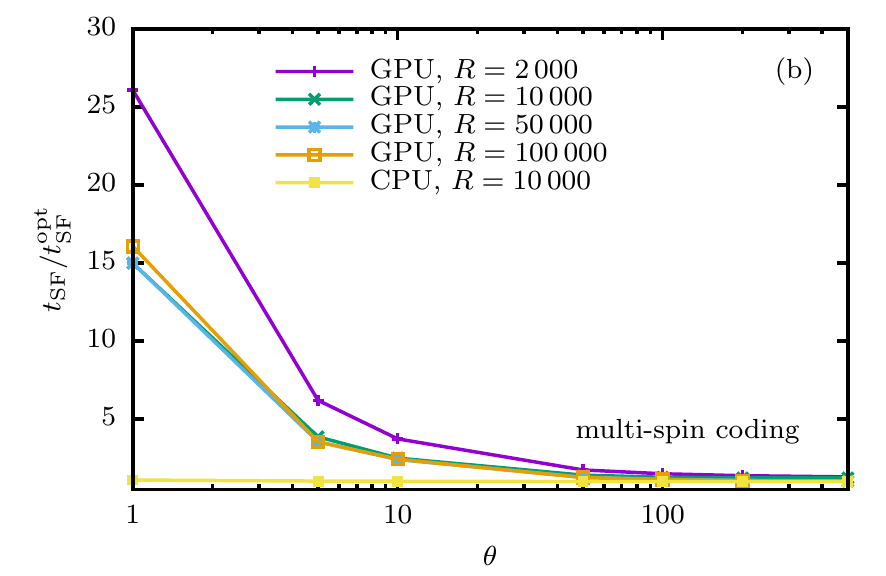}
    \caption{%
      (a) Time per spin flip $t_\mathrm{SF}$ of the single-spin coded GPU and CPU codes as a function
      of the number $\theta$ of equilibration sweeps relative to the time
      $t_\mathrm{SF}^\mathrm{opt}$ achieved in the fastest case considered, namely
      for $\theta=500$. The different lines show different population sizes. (b) The
      same comparison for the multi-spin coded GPU code. The reference line for CPU
      is again for the single-spin coded algorithm. All data are for $L=128$.
    }
    \label{fig:performance}
  \end{center}
\end {figure}

A method that provides high performance without compromising the statistical quality
of data can be constructed by combining the underlying RNG with a particularly fast
in-line generator used to supply the $p$ random numbers used to flip the spins
stored in the same word. For this purpose, we use a simple linear-congruential
generator (LCG) of the form
\[
r_{n+1} = A r_{n}+C \mod 2^{32},
\]
with $A=1\,664\,525$ and $C=1\,013\,904\,223$ \cite{numrec}. For each call to the
spin-updating kernel and each word of $p$ spins, this generator is seeded by a call
to the underlying, high-quality generator (Philox). Although LCG generators are no
longer recommended for general purpose applications in simulations (see, e.g.,
Ref.~\cite{manssen:12} and references therein), we believe that this does not cause
any problems in the present context as the LCG is only used to multiply a sample of
the underlying RNG and additionally the resampling is done with the base RNG
alone. Empirical testing confirms this assumption as no biases or increases in
statistical fluctuations are observed. The corresponding results shown in
Fig.~\ref{fig:msc}(b) and (c) reveal that the relative error for this approach is
identical to that of simulations using $p$ samples from the base RNG. As the data in
the last section of Table~\ref{tab:msc} illustrate, the performance of this combined
approach is excellent, providing an about 10-fold speed-up of the simulations with
MSC and $p=32$ as compared to the simulations with SSC (both running on GPU).

\section{Performance}
\label{sec:performance}

\begin{table}[tb!]
  \centering
  \caption{
    Peak performance of the CPU and GPU PA implementations in units of the total run
    time divided by the total number of spin flips performed, $t_\mathrm{SF}$, for
    different system sizes. The best GPU performance is achieved for large $\theta$,
    and here  $\theta=500$ was chosen for a population of $R=50\,000$ replicas. The
    speedups for the SSC and MSC GPU codes are relative to the CPU results. GPU
    performance data are for the  Tesla K80. The CPU code was benchmarked on an Intel
    Xeon E5-2683 v4 CPU running at 2.1 GHz.
  }
  \label{tab:peak-performance}
  \begin{tabular}{rccccc}
    \hline
    & CPU & \multicolumn{4}{c}{GPU} \\
    \cline{3-6}\\[-2ex]
    & & \multicolumn{2}{c}{SSC} & \multicolumn{2}{c}{MSC} \\
    $L$ & $t_\mathrm{SF}$ [ns] & $t_\mathrm{SF}$ [ns] & speedup &
    $t_\mathrm{SF}$ [ns] & speedup \\
    \hline
    16 & 23.1 & 0.094 & 246 & 0.0096 & 2406 \\
    32 & 22.9 & 0.092 & 249 & 0.0095 & 2410 \\
    64 & 22.6 & 0.092 & 246 & 0.0098 & 2306 \\
    128 & 22.6 & 0.097 & 233 & 0.0098 & 2306 \\
    256 & 22.5 & 0.098 & 230 & 0.0099 & 2273 \\
    \hline
  \end{tabular}
\end{table}

In order to compare performance across different choices of the algorithmic
parameters $R$, $\theta$ and $\Delta\beta$, we normalize the time for a full PA run
by the total number of spin flips performed,
\begin{equation}
  \label{eq:spinflip}
  t_\mathrm{SF} = \frac{t_\mathrm{run}}{L^2 \theta \sum_{i=1}^{N_\beta} R_i},
\end{equation}
where $N_\beta$ denotes the number of temperature steps. We compare the GPU codes
proposed here to our reference CPU implementation which is an optimized scalar
program, so only uses one core. Instead of discussing the performance of a range of
different CPUs and GPUs, we here restrict ourselves to the GPUs and CPUs available in
an HPC cluster machine recently installed at the home institution of one of us
(Coventry University), which are Intel Xeon E5-2683 v4 CPUs and NVIDIA Tesla K80 GPU
cards, which should be fairly representative of present-day HPC cluster
configurations. The K80 is a double card, of which only one card is actually used for
the measurements at a time, featuring 2880 cores and 12GB of RAM. We note that we
tested a variant of the CPU code parallelized on a single CPU using OpenMP and found
close to perfect parallel scaling efficiency. A comparison to a parallel code for a
particular CPU can therefore be achieved by dividing the speedup factors quoted here
by the number of cores in the processor used.

In general, the best performance in terms of the metric \eqref{eq:spinflip} is
achieved when minimizing the frequency of resampling steps, such that practically all
time is invested in flipping spins. First focusing on this optimal case, we collect
in Table \ref{tab:peak-performance} the times $t_\mathrm{SF}$ for $\theta=500$ and a
population size $R=50\,000$ for different system sizes. It is seen that in the range
$16\le L\le 256$ considered the performance of all three codes is almost independent
of system size. For the GPU codes this is an indication that through the replica
parallelism and the additional spin-parallelism there is enough parallel work to
saturate the device already for moderate system sizes. The single-spin coded GPU code
is found to be at least 230 times faster than the CPU case. The multi-spin coding
with $p=32$ and the additional combination with an LCG generator yields a further
factor of 10, resulting in a total peak speedup of the MSC code of about 2300 as
compared to the scalar program.

\begin{table}[tb]
  \centering
  \caption{Times $t_\mathrm{SF}$ per spin flip (in ns) for SSC and MSC GPU codes run on the Tesla K80
    GPU for a $L=128$ system.}
  \label{tab:times}
  \begin{tabular}{crllll}
    \hline
    \multicolumn{2}{c}{} & \multicolumn{4}{c}{$R$} \\
    & & 2\,000 & 10\,000 & 50\,000 & 100\,000\\ \hline
   \multicolumn{2}{c}{} & \multicolumn{4}{c}{single-spin coding (SSC)} \\ \hline
    &  1   &  0.229 &  0.213 &  0.213  &   0.213 \\
    &  5   &  0.123 &  0.119 &  0.119  &   0.120 \\
    & 10  &  0.111 &  0.108 &  0.107  &   0.109 \\
    $\theta$ &  50  &  0.101 &  0.0994 &  0.0985  &   0.0987 \\
    &  100 &  0.0998 & 0.0976 &  0.0977 &  0.0975 \\
    &  200 &  0.0997 & 0.0972 &  0.0970 &  0.0970 \\
    &  500 &  0.0991 & 0.0971 &  0.0969 &  0.0969 \\
    \hline
    \multicolumn{2}{c}{} & \multicolumn{4}{c}{multi-spin coding (MSC)} \\ \hline
    &  1   & 0.2504 & 0.1439 &  0.1440 &  0.1543 \\
    &  5   & 0.0596 & 0.0372 &  0.0341 &  0.0341 \\
    &  10  & 0.0359 & 0.0240 &  0.0232 &  0.0234 \\
    $\theta$ &  50  & 0.0168 & 0.0136 &  0.0123 &  0.0121 \\
    &  100 & 0.0144 & 0.0123 &  0.0110 &  0.0108 \\
    &  200 & 0.0132 & 0.0119 &  0.0103 &  0.0101 \\
    &  500 & 0.0125 & 0.0118 &  0.0098 &  0.0097 \\
    \hline
  \end{tabular}
\end{table}

While the performance of the CPU code is almost independent of $\theta$ and $R$ in
the ranges studied here, the speed of the GPU codes varies significantly with these
parameters, in particular with $\theta$. This is illustrated in
Fig.~\ref{fig:performance}, while the corresponding data is collected in
Table~\ref{tab:times}. For the SSC program, there is almost no dependence on $R$ in
the range $2\,000\le R\le 100\,000$ shown here, but for small values of $\theta$ the
times per spin flip increase by up to a factor of 2.4 as compared to the optimal
case, cf.\ Fig.~\ref{fig:performance}(a). Hence, for the extreme case of $\theta=1$,
the speedup reduces to a factor of 100. For $\theta=10$, on the other hand, which was
typically used in previous applications of the PA algorithm \cite{wang:14,wang:15},
the performance is only about 10\% below the optimum and the speedup is still
approximately 200. For the multi-spin coded program, this effect becomes even much
more pronounced as the time per spin flip is reduced by a factor of 10, but the time
taken for the resampling is not improved at the same rate. Additionally, the
proportion of time taken for the sampling of observables increases significantly. As
a consequence, there is a rather strong $\theta$ dependence of the performance of the
MSC version, which is 15--20 times slower for $\theta=1$ than for $\theta=500$, see
the data shown in Fig.~\ref{fig:performance}(b). In this extreme case, the speedup is
reduced to about 150, almost comparable to the performance of the SSC program. For
the choice $\theta=10$, on the other hand, the MSC code still performs at 950 times
the CPU code's speed, see also the data collected in the lower part of Table
\ref{tab:times}.

\begin{table}[tb]
  \centering
  \caption{The fraction of the total computing time spent in the spin-flip kernel
    \texttt{checkKerALL()} of the SSC and MSC GPU codes for a $L=128$ system
    simulated on the Tesla K80 GPU.}
  \label{tab:fraction}
  \begin{tabular}{crrrrr}
    \hline
    \multicolumn{2}{c}{} & \multicolumn{4}{c}{$R$} \\
                         & & 2\,000 & 10\,000 & 50\,000 & 100\,000\\ \hline
     \multicolumn{2}{c}{} & \multicolumn{4}{c}{single-spin coding (SSC)} \\ \hline
                         &    1 & 39.8\%    &  41.8\%    &    42.2\%   & 42.2\%    \\
                         &    5 & 76.4\%    &  77.9\%    &    78.3\%   & 78.3\%    \\ 
                         &   10 & 86.6\%    &  87.6\%    &    88.0\%   & 87.9\%    \\
    $\theta$             &   50 & 97.0\%    &  97.3\%    &    97.4\%   & 97.3\%    \\
                         &  100 & 98.5\%    &  98.6\%    &    98.7\%   & 98.6\%    \\
                         &  200 & 99.2\%    &  99.3\%    &    99.3\%   & 99.3\%    \\
                         &  500 & 99.7\%    &  99.7\%    &    99.7\%   & 99.7\%    \\
    \hline
    \multicolumn{2}{c}{} & \multicolumn{4}{c}{multi-spin coding (MSC)} \\ \hline
                         &  1   & 5.3\%  &  6.6\%  &  5.9\%   &   5.4\%     \\
                         &  5   & 20.6\%  & 28.7\%  & 25.1\%   &  24.4\%     \\
                         &  10  & 33.9\%  & 44.5\%  & 37.3\%   &  35.9\%     \\
    $\theta$ &  50  & 71.8\%  & 80.0\%  & 75.5\%   &  76.1\%     \\
                         &  100 &  83.6\%  & 89.0\%  & 86.7\%   &  86.3\%     \\
                         &  200 & 91.0\%  & 94.1\%  & 93.2\%   &  92.9\%     \\
                         &  500 & 96.2\%  & 97.8\%  & 97.3\%   &  97.1\%     \\
    \hline
  \end{tabular}
\end{table}

When discussing the optimization of the GPU code above, we stated that most of the
time in PA is spent on spin flips. To see whether this is indeed the case, we
calculated the fraction of the total run time spent in the spin updating kernel
\texttt{checkKerALL()}. The corresponding data are collected in Table
\ref{tab:fraction}. For the SSC program, the percentage of time spent updating spins
is indeed generally high, and exceeding 85\% for $\theta\ge 10$. On the other hand,
for the minimal $\theta=1$ it drops to about 40\%. For the MSC version, the relative
cost of resampling and measurements of the energy and magnetization is much more
significant, and a fraction of 85\% of time for spin flips is only reached for
$\theta = 100$, see the lower part of Table \ref{tab:fraction}. The CPU code, in
contrast, spends 90\% of time on flipping spins even for $\theta = 1$. These
differences are a consequence of the additional overhead resulting from the parallel
reductions for calculating $Q_i$ and the resampling factors $\hat{\tau}_i$, as well
as copying replicas and measuring observable values. For the MSC version of the code
there is the additional complication that for the energy and magetization calculation
the individual spins need to be unpacked from the bit-coded words, which is quite
costly.

Nevertheless, it is crucial to move as much of the calculation as possible onto GPU
instead of possibly using a hybrid approach. This is illustrated in Table
\ref{tab:copy}, where we compare the performance in terms of $t_\mathrm{SF}$ for a
variant that does the spin flips on GPU, but transfers the population of replicas
back to CPU for the implementation of the resampling process. At least for small
values of $\theta$, this hybrid version is significantly less performant. On the
other hand, it is conceptually simpler as it does not make use of parallel reductions
etc., so users could consider this simpler approach for simulations where a large
$\theta$ is used.

\section{Conclusion}
\label{sec:conclusions}

We have presented an efficient implementation of the population annealing algorithm
on GPUs, using the 2D Ising model as a benchmark problem. The code takes into account
a range of fundamental optimization heuristics for GPU computing, including the
principles of latency hiding and memory coalescence and thus achieves peak speedups
of more than 200 times above a reference serial implementation on CPU. To create
sufficient parallel work for the GPU devices it turns out to be useful to combine the
replica-level parallelism with an additional domain decomposition, thus exploiting
also spin-level parallelism. We also provide here a multi-spin coded version of the
program that yields peak performances of more than 2000 times that of the serial,
single-spin coded variant. After completing the main work of this paper, we had the
opportunity of performing some test runs of our code on a GeForce GTX 1080 GPU. As
this is a Pascal generation card, it promises some speedups compared to the Kepler
board K80. For $L=64$, $R=50\,000$ and $\theta=500$ we find the peak performance to
be 36 ps for SSC and 4 ps for MSC, such that the maximal speedup factors against the
serial CPU code increase to 625 and 5650, respectively.

\begin{table}[bt]
  \centering
  \caption{%
    Time $t_\mathrm{SF}$ per spin flip for a hybrid version of the SSC GPU code,
    where at each temperature step the full population is copied between GPU and CPU to perform
    the resampling as compared to the
    fully GPU-embedded standard version proposed here. While for large values of $\theta$
    both versions show similar performance, for small $\theta$ the copying slows down the
    hybrid version significantly, a fact that is consistent with the general observation
    \cite{hwu:11} that fully GPU enabled code is almost always preferable over hybrid
    solutions. Simulations are for $L=128$ and $R=50\,000$ on the NVIDIA Tesla K80.%
  }
  \label{tab:copy}
  \begin{tabular}{crrrrrr}
    \hline
    & \multicolumn{6}{c}{$t_\mathrm{SF}$ [ns]} \\
      \hline
     $\theta$   &      1 & 5 & 10 & 50 & 100 & 500 \\ \hline
    standard & 0.213 & 0.119 & 0.107 & 0.099 & 0.098 & 0.097 \\
    hybrid & 1.424 & 0.361 & 0.232 & 0.126 & 0.112 & 0.101 \\
   \hline
  \end{tabular}
\end{table}

While we provide code for the 2D Ising ferromagnet, we hope it to be used as a
template for the simulation also of further problems with the same
algorithm. Generalizations for models on different lattices in various dimensions,
the case of different couplings including spin glasses and random-field systems as
well as more general spin systems such as Potts \cite{barash:17} or O($n$) models are
straightforward. Applications to off-lattice systems \cite{callaham:17} are also
straightforward conceptually, although the optimization of the code in such cases
might be a little bit more difficult.

Large populations can be simulated on standard GPUs. For the present implementation,
for $L=64$ it is possible on the K80 GPUs to simulate $R=1.5\times 10^6$ replicas
using the SSC variant and $R=1.2\times 10^7$ for the MSC version. It is worthwhile
noting that one can combine simulations to effectively achieve the precision expected
from a single run with the combined population size by using the weighted averaging
scheme as discussed above in Sec.~\ref{sec:weighted}
\cite{machta:10a,weigel:16}. Additionally, it should be possible to combine GPU
parallelization with MPI and run very large populations on a cluster of GPU enabled
nodes, and we expect population annealing to show excellent scaling properties for
such setups.

\section{Acknowledgments}

The work of L.B., M.B.\ and L.S.\ is supported by the grant 14-21-00158 from the
Russian Science Foundation. M.B. was also supported by the Scientific Grant Agency of
the Ministry of Education of the Slovak Republic (Grant No. 1/0331/15). The authors
acknowledge support from the European Commission through the IRSES network DIONICOS
under Contract No. PIRSES-GA-2013-612707. The simulations were performed on the HPC
facilities of Coventry University and the Science Center in Chernogolovka. M.W.\
acknowledges fruitful and pleasant discussions with Jon Machta and Helmut Katzgraber
on the subject of population annealing.

%\bibliography{citeulike_nourl_noissn}

\end{document}